\documentclass[12pt
]{article}%
\usepackage[font=footnotesize]{caption}
\usepackage[export]{adjustbox}
\usepackage{float}
\usepackage{placeins}
\usepackage{dsfont}
\usepackage[caption = false]{subfig}
\usepackage{graphicx}
\usepackage[update]{epstopdf}
\usepackage{rotating}
\usepackage[mathscr]{euscript}
\usepackage{enumerate}
\usepackage{dcolumn}
\usepackage{bbm}
\usepackage{soul}
\usepackage{lscape}
\usepackage{longtable}
\usepackage{multirow}
\usepackage{setspace}
\usepackage{threeparttable}
\usepackage[centertags,sumlimits, intlimits, namelimits,]{amsmath}
\usepackage{amsfonts}
\usepackage[english]{babel}
\usepackage{amssymb}
\usepackage{appendix}
\usepackage{fancyhdr}
\usepackage{booktabs}
\usepackage{natbib}
\usepackage[applemac]{inputenc}
\usepackage{url}
\usepackage{longtable}
\usepackage[small]{titlesec}
\usepackage{geometry}
\usepackage{eurosym}
\usepackage{comment}
\usepackage{authblk}
\usepackage[normalem]{ulem}
\usepackage{dcolumn}
\usepackage{placeins}

\newcommand{\mc}{\multicolumn{1}{c}}

\usepackage{rotating}

\usepackage[usenames,dvipsnames]{xcolor}
\usepackage{hyperref}

\usepackage{xcolor}
\definecolor{mycolor}{rgb}{0.55, 0.0, 0.0}

\hypersetup{
	pdftoolbar=true,        
	pdfmenubar=true,        
	pdffitwindow=false,     
	pdfstartview={FitH},    
	pdftitle={Intertemporal Consumption and Debt Aversion: A Replication and Extension},    
	pdfauthor={Steffen Ahrens, Ciril Bosch-Rosa, Thomas Meissner},     
	pdfsubject={Replication Meissner (2016)},   
	pdfkeywords={Replication}, 
	pdfnewwindow=true,      
	colorlinks=true,       
	linkcolor=mycolor,          
	citecolor=mycolor,        
	filecolor=magenta,      
	urlcolor=mycolor,           
	pdfpagemode = UseNone,
}
\usepackage[hyperpageref]{backref}

\usepackage{pdfpages}

\renewcommand*{\backref}[1]{}
\renewcommand*{\backrefalt}[4]{
	\ifcase #1
	Not cited
	\or
	Cited on page~#2.
	\else
	Cited on pages~#2.
	\fi}

\def\sym#1{\ifmmode^{#1}\else\(^{#1}\)\fi}
\renewcommand\abstract{{\noindent\bfseries Abstract\\[1.5ex]}}
\def\keywordname{{\bfseries Keywords}}%
\def\keywords#1{\par\addvspace\medskipamount{\rightskip=0pt plus1cm
		\def\and{\ifhmode\unskip\nobreak\fi\ $\cdot$
		}\noindent\keywordname\enspace\ignorespaces#1\par}}
\def\JELname{{\bfseries JEL Classification}\enspace}
\def\JEL#1{\par\addvspace\medskipamount{\rightskip=0pt plus1cm
		\def\and{\ifhmode\unskip\nobreak\fi\ $\cdot$
		}\noindent\JELname\ignorespaces#1\par}}
\def\ackname{Acknowledgments}%
\def\acknowledgement{\par\addvspace{17pt}\small\rmfamily
	\trivlist\if!\ackname!\item[]\else
	\item[\hskip\labelsep
	{\bfseries\ackname}]\fi}

\renewcommand{\tablename}{TABLE}
\renewcommand\thetable{\arabic{table}}
\def\fnum@table{\tablename\nobreakspace\thetable}
\newcolumntype{d}[1]{D{.}{.}{#1}}

\begin{document}

	\title{\vspace{-0mm} 
	Intertemporal Consumption and Debt Aversion: A Replication and Extension\footnote{
			Corresponding author: \href{mailto:steffen.ahrens.econ@gmail.com}{steffen.ahrens.econ@gmail.com}. We would like to thank Frank Heinemann, the editor and two anonymous referees for their comments. Steffen Ahrens and Ciril Bosch-Rosa gratefully acknowledge financial support from the Deutsche Forschungsgemeinschaft (DFG) through the CRC TRR 190 (project number 280092119, ``Rationality and Competition''). Replication files can be found via OSF: \url{https://osf.io/9cgkf/?view_only=8a5997b9a26d42c8b8a529bd95008d9e}}
	}
\author[1]{Steffen Ahrens}
\author[2]{Ciril Bosch-Rosa}
\author[3]{Thomas Meissner}
\affil[1]{Chair of Macroeconomics, Freie Universit\"{a}t Berlin}
\affil[2]{Chair of Macroeconomics, Technische Universit\"{a}t Berlin}
\affil[3]{Department of Microeconomics and Public Economics, Maastricht University}

	\maketitle
	\bigskip
	\bigskip
	
	\renewcommand{\baselinestretch}{1} \normalsize
	\begin{abstract}

		We replicate \cite{Meissner2016}, where debt aversion was reported for the first time in an intertemporal consumption and saving problem. While \cite{Meissner2016} uses a German sample, our participants are US undergraduate students. All of the original study's main findings replicate with similar effect sizes. Additionally, we extend the original analysis by introducing a new individual index of debt aversion, which we use to compare debt aversion across countries. Interestingly, we find no significant differences in debt aversion between the original German and the new US sample. We then test whether debt aversion correlates with individual characteristics such as gender, cognitive reflection ability, and risk aversion. Overall, this paper confirms the importance of debt aversion in intertemporal consumption and saving problems and validates the approach of \cite{Meissner2016}.

		\JEL{C91 \and D84 \and G11 \and G41}
	\end{abstract}
	\renewcommand{\baselinestretch}{1.45} \normalsize
	\setlength{\footnotesep}{0.75\baselineskip}
	\newpage

\section{Introduction}

Debt is a powerful tool to allocate resources over time. Used appropriately, it increases welfare and fosters growth \citep{cecchetti2011}. Yet, many people show an aversion to debt with far-reaching consequences for individual welfare and economic growth. For instance, debt averse entrepreneurs might pass on profitable investment opportunities \citep{paaso2021entrepreneur}, debt averse households might waive profitable retrofit investments \citep{schleich2021adoption}, and debt averse high school students might forego a college or university degree \citep{ callenderjackson2005,boatman2017,callendermason2017}.

In a recent laboratory experiment with German undergraduates, \cite{Meissner2016} studies the role of debt in an intertemporal consumption and saving problem.\footnote{For an extensive survey of laboratory experiments on dynamic stochastic optimization problems see \cite{Duffy2016}.} According to theory, agents optimally allocate their expected lifetime income over time, saving when income is high and borrowing when income is low \citep[e.g.,][]{fisher1930, Friedman1957, Modigliani1986}. By contrast, the experimental results of \cite{Meissner2016} show that participants generally fail to solve such intertemporal optimization problems. Furthermore, participants are less willing to borrow than they are willing to save to smooth consumption. The author interprets this asymmetry as an indication of debt aversion.

This paper is an exact replication in the sense of \cite{chen2021best} of the experiment by \cite{Meissner2016}.\footnote{While the author of the original experiment is an author of the present paper, he was not directly involved in running the new experimental sessions.} There are several reasons to replicate this study. First, debt aversion is a relevant problem that has not yet received much attention in the dynamic optimization literature (see, e.g., \cite{Duffy2016}). Replicating existing work lends credibility to the limited existing results. Second, the task in the original experiment is complex, so reproducing the original results will help establish a reliable experimental design to study debt aversion. Third, \cite{Meissner2016} uses a sample of the student population in Germany, a country which - by international standards - is known for moderate levels of household debt \citep[e.g.,][]{Christelisetal2021}, an excessive reliance on cash payments  \citep[e.g.,][]{kalckenreuthetal2014,Bagnalletal2016}, and low tuition fees for higher education \citep[e.g.,][]{OCED2021}, which imply low levels of student debt. Therefore, the observed debt aversion in \cite{Meissner2016} could be specific to populations without previous experience acquiring debt, or even specific to Germany, which is known for its cultural abhorrence of debt. As Nietzsche notes, in German debt is spelled as ``Schuld,'' which means both ``debt'' and ``guilt,'' to argue that ``debt'' \textit{with oneself} is the source of guilt and bad conscience \citep{nietzsche2021genealogie}.

It is well-known that culture matters in experimental settings \citep{henrich2001search, chen2021best}. Against this background, we use a population composed of undergraduate students at the University of Illinois at Urbana-Champaign (UIUC) to test the robustness of the results of  \cite{Meissner2016}.  The US is known for having a more tolerant view of debt \citep{calder2009financing} and for encouraging it through its institutions \citep{garon2011beyond}. Therefore, as is common in the United States, students at UIUC incur student debt to pay for tuition fees and other expenses during their studies. The US Department of Education reports an average annual cost of studying at UIUC of \$15,880 and a median total debt after graduation between \$15,000 and \$26,000 depending on the field of study.\footnote{This cost includes tuition, living costs, books and supplies, and fees minus the average grants and scholarships for federal financial aid recipients. 
}
Therefore, it is safe to assume that the student body at UIUC is less restrictive about acquiring debt and has more homegrown experience acquiring it compared to German students. 

Furthermore, we extend the original analysis of \cite{Meissner2016} by developing an index of debt aversion that allows us to compare debt aversion of students in the original sample of \cite{Meissner2016} to the students from UIUC. Additionally, we collect information on participants' gender, risk aversion, and cognitive reflection ability, as measured by the Cognitive Reflection Test \citep[CRT,][]{Frederick2005}. We are especially interested in the cognitive reflection of participants, as it is a strong determinant of financial behavior both in and outside the laboratory (see \cite{gomes2021household} and \cite{bosch2021cognitive} for an overview of results in the field and the lab, respectively).\footnote{There is a vivid debate in the literature on whether the CRT \textit{is} a measure of cognitive ability or an independent rationality factor \citep[e.g.,][]{Frederick2005, campitelli2014does, pennycook2016cognitive}.  On the one hand, \cite{toplak2011cognitive} and \cite{toplak2014assessing} argue that CRT is an indicator of rational thinking performance that is independent and separable from cognitive ability. On the other hand, in a recent meta-analysis \cite{otero2022cognitive} conclude that ability to solve the CRT cannot be interpreted as an independent cognitive factor, but rather as a combination of cognitive ability and numerical ability. To avoid any confusion, for the remainder of the paper we opt to refer to what the   CRT measures as ``cognitive reflection.''} 


Our results show that the findings of \cite{Meissner2016} replicate. Participants fail to smooth consumption optimally and are disproportionately more reluctant to smooth consumption via debt compared to savings. Moreover, the effect sizes are similar and there appears to be no difference in the degree of debt aversion between the two samples. Testing for correlation with individual characteristics, we find no evidence that risk aversion or gender correlate with debt aversion. However, we find some weak evidence suggesting that cognitive reflection ability could be negatively correlated with debt aversion. 




To our knowledge, this is the first intertemporal consumption and saving experiment to compare the behavior between an American and a European sample. Moreover, existing literature on debt aversion is scant, and we are not aware of any direct intercultural comparisons. However, some recent related empirical evidence exists: \cite{hundtofte2019credit} test whether individuals in Iceland and the US use short term credit to smooth consumption when they experience a transitory negative income shock. They find that individuals from neither Iceland nor the US use short term credit to smooth consumption, but rather adjust consumption downwards. This is in line with observed behavior in our experiment, where participants are also reluctant to borrow to smooth consumption. 

The remainder of the paper is organized as follows. Section \ref{sec:design} presents the experimental design, Section \ref{sec:results} reports the results of the replicated experiment and how personal characteristics correlate with the new debt aversion index. Finally, Section \ref{sec:conclusion} concludes.

\section{Experimental Design}\label{sec:design}
The design of the experiment is identical to \cite{Meissner2016} and implements a simple life-cycle model of consumption. In each period of a life-cycle $(t=1,\dots,20)$, participants choose how much of their wealth $(w_t)$ to consume $(c_t)$ and how much to save $(a_t)$. Savings can be positive or negative, where negative savings are referred to as debt. We abstract from any interest payments on savings or debt and there is no discounting. Each period, participants are provided with an exogenous income $(y_t)$, which follows a trend stationary stochastic process. Consequently, wealth in period $t$ is defined as $w_t = y_t + a_{t-1}$. In the initial period of a life-cycle, participants start with zero savings $(a_0=0)$ and in the final period of the life cycle all wealth has to be consumed as saving is not possible ($a_{20}=0$). Taken together, the latter two restrictions imply that life-cycle consumption must be equal to life-cycle income, i.e.  $\sum_{t=1}^{20}c_t=\sum_{t=1}^{20}y_t$.

Consumption decisions are incentivized using a time-separable CARA utility function of the form $u(c_t)=250\left( 1-e^{-\theta c_t}\right)$, where $\theta$ denotes the parameter of absolute risk aversion, which we set equal to $\theta=0.02$ as in \cite{Meissner2016}.
The participant's objective is to choose a stream of consumption that maximizes her life-cycle utility. Therefore, in any period $t$, the decision problem of participants is given by:
\begin{eqnarray}
  \max_{c_t} E_t \sum_{t=\tau}^T u(c_t),\\
  c_t + a_t = w_t,\\
  w_t = y_t + a_{t-1},\\
  a_0=0, a_T=0.
\end{eqnarray}

Given CARA utility, \cite{MeissnerRostamAfschar2017} show that for any income process $y_t=y_0+st+\varepsilon_t$, where $P\left( \varepsilon_t = \sigma_\varepsilon \right)=P\left( \varepsilon_t =- \sigma_\varepsilon \right)=0.5,\, \forall t,$ period-$t$ optimal consumption is given by
\begin{eqnarray}
  c^\ast_t(w_t) = \frac{1}{T-t+1}\left[w_t+\zeta_t - \Gamma_t (\theta \sigma_\varepsilon)\right], \label{eq:opt_solution_1}\\
  \zeta_t = (T-t)\left( y_0 + s\left( \frac{T+t+1}{2} \right) \right),\label{eq:opt_solution_2}\\
  \Gamma_t(\theta\sigma_\varepsilon)= \sum_{j=0}^{T-t}\sum_{i=1}^j \log \cosh \left( \frac{\theta \sigma_\varepsilon}{T-t+1-i} \right),\label{eq:opt_solution_3}
\end{eqnarray}
where $\zeta_t$ is the expected life-time income $\zeta_t=E_t\left[\sum_{j=1}^{T-t}y_{t+j} \right] $ and $\Gamma_t(\theta\sigma_\varepsilon)$ are precautionary savings. Equations \eqref{eq:opt_solution_1} to \eqref{eq:opt_solution_3} imply a smooth consumption path over the life-cycle for the given income process specified above.

\begin{figure}[t]
\caption{Example increasing income stream (dashed line), decreasing income stream (dotted line), and optimal consumption (solid line) \label{fig:income_and_opt_consumption}}
\includegraphics[width=\textwidth]{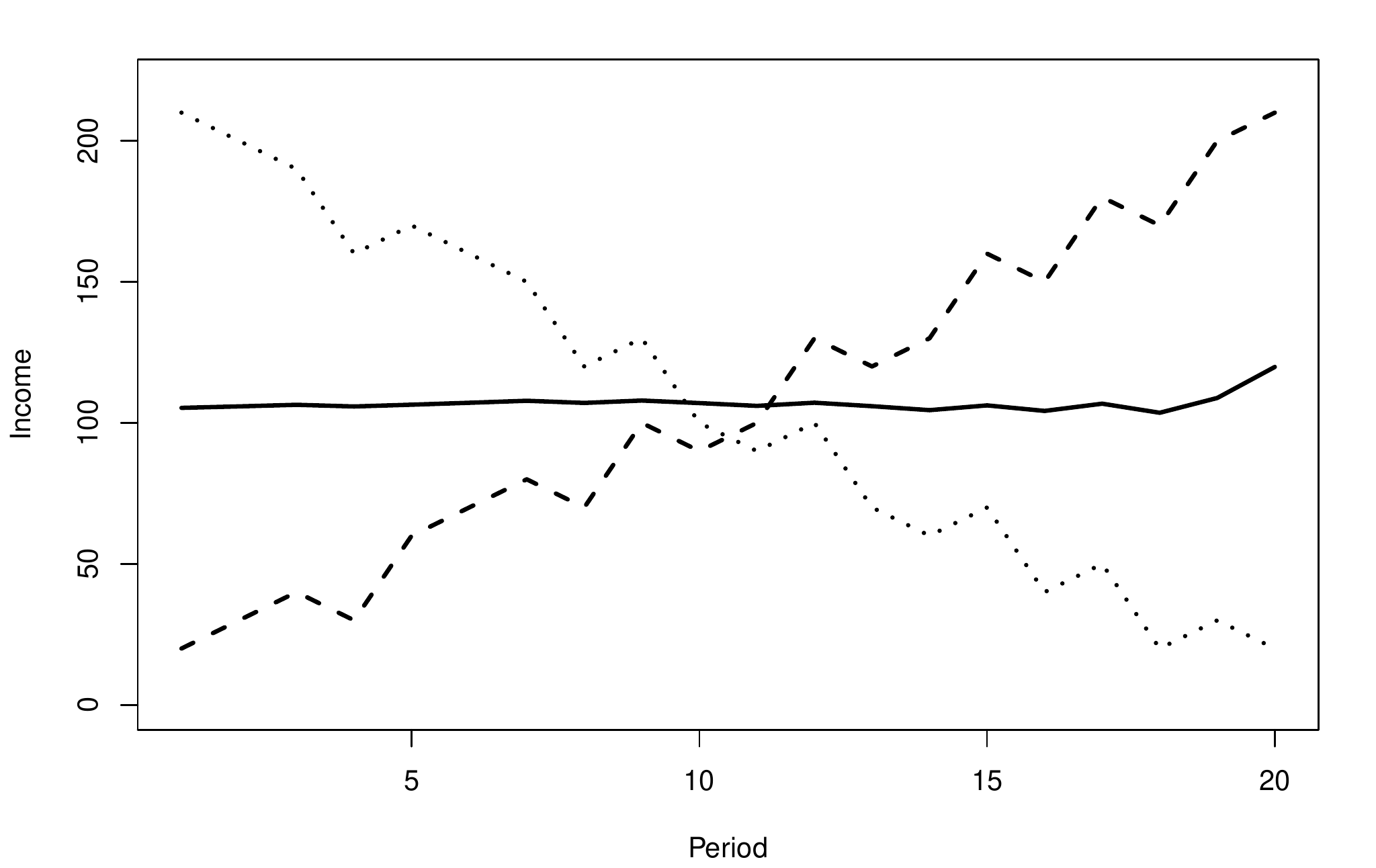}

\end{figure}

The treatments in this experiment differ with respect to the income process. In the \textit{borrowing treatment}, participants face an income process $y_t^B=10t+\varepsilon_t$, which increases over the life cycle. To smooth consumption, participants have to borrow early on in their life-cycle and repay their debt from high income later in the life-cycle. In the \textit{saving treatment}, participants face a decreasing income process given by $y_t^S=210-10t+\varepsilon_t$. Here participants have to save early in the life-cycle and then live of their savings later on. In each period, the shock $\varepsilon_t$ takes the value of $+10$ with 50\% probability and the value of $-10$ with 50\% probability.
Given the same shock sequence, equations \eqref{eq:opt_solution_1} to \eqref{eq:opt_solution_3} imply the same optimal consumption path for both the increasing and the decreasing income process. Figure \ref{fig:income_and_opt_consumption} provides an exemplary increasing (dashed line) and decreasing (dotted line) income processes for a given shock sequence and the associated optimal consumption path (solid line).\footnote{The shock sequence varies between rounds, but is the same for all sessions and participants.}

To assess learning effects and to add a within-subject dimension, each participant plays three rounds of the borrowing treatment and three rounds of the savings treatment.

\subsection{Experimental Procedures}

Each session consisted of six rounds, each 20 periods long. In the Borrowing First (BF) sessions, participants first played the \textit{borrowing treatment} for three rounds followed by three rounds of the \textit{saving treatment}. In the Saving First (SF) sessions, the order of the treatments was inverted. While participants knew that the session had six rounds, the specific instructions for each type of income process were read immediately before the start of each three-round sequence. As this experiment is an exact replication, we refer the reader to \cite{Meissner2016} for further details on the experimental procedures.

After the experiment, participants were asked to fill out a questionnaire which contained a hypothetical multiple price list to assess individual risk aversion, the cognitive reflection test (CRT), and some individual characteristics, such as gender, field of study, and nationality. We also asked participants if they had previously seen the CRT questions.\footnote{Around 15\% of participants reported to have seen the CRT questions previous to our experiment. Interestingly, participants who report to know the CRT do not score significantly higher (Mann-Whitney U Test, $p=0.69$).} The instructions of the experiment and the questionnaire can be found in Appendix~\ref{sec:instructions}. 

\section{Results}\label{sec:results}

The experiment was conducted during the fall of 2016 at the University of Illinois at Urbana-Champaign and the experimental software was written in z-Tree \citep{Fischbacher2007}. A total of 91 participants took part in the experiment, 44 in the Borrowing First sessions and 47 in the Saving First sessions. Most of the participants were undergraduate students in the field of business, engineering, and economics, similar to \cite{Meissner2016}. Table \ref{tab:summarystats} contains summary statistics on CRT score, gender and risk aversion of our sample. Each session lasted around 60 minutes and participants earned \$19.12 on average. The minimum payment was \$5.50. 

\begin{table}[t!]
\centering
\caption{Summary statistics \label{tab:summarystats}}
\begin{tabular}{l c c c  c c}\toprule
\multicolumn{1}{c}{Variable} & Obs & Mean & Std. Dev.
  & P5 & P95  \\ \midrule
CRT score & 91 & 1.967 & 1.069  & 0 & 3 \\
Female & 90 & 0.389 & 0.49  & 0 & 1 \\
Risk aversion & 85 & 6.518 & 3.8  & 0 & 14 \\
\bottomrule\end{tabular}

\begin{tablenotes}
\footnotesize
        \item[a] \emph{Notes:} \emph{CRT score} is the number of correct answers in the CRT. \emph{Female} takes the value one for female participants and zero otherwise. \emph{Risk aversion} contains the number of safe options in a multiple price list.
    \end{tablenotes}
\end{table}

\subsection{Consumption Choices}
\begin{figure}
\caption{Median and mean consumption \label{fig:mean_consumption}}
\hspace*{6.8cm}\textbf{Borrowing First}\\
\includegraphics[width=.95\textwidth]{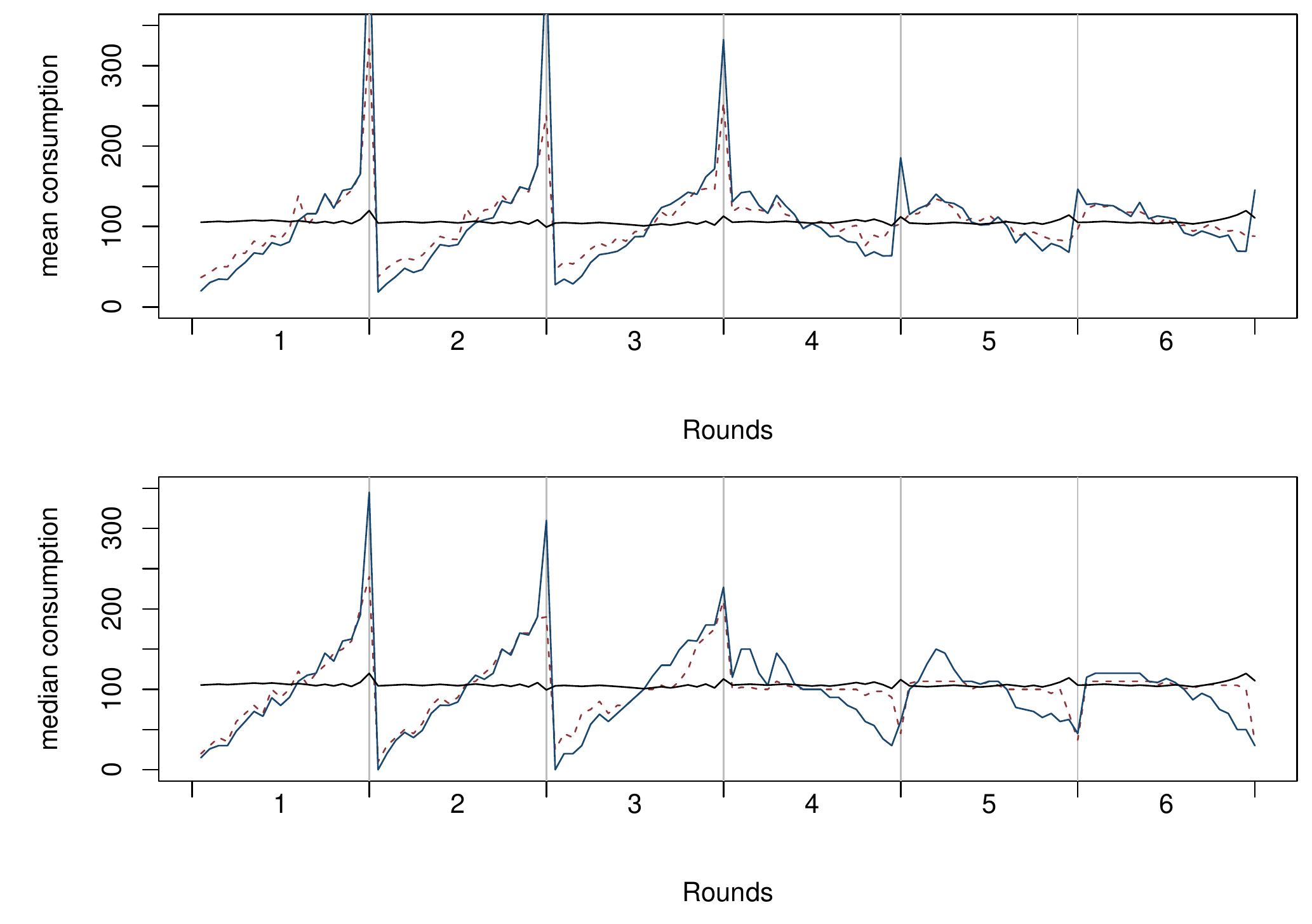}\\
\hspace*{7cm}\textbf{Saving First}\\
\includegraphics[width=.95\textwidth]{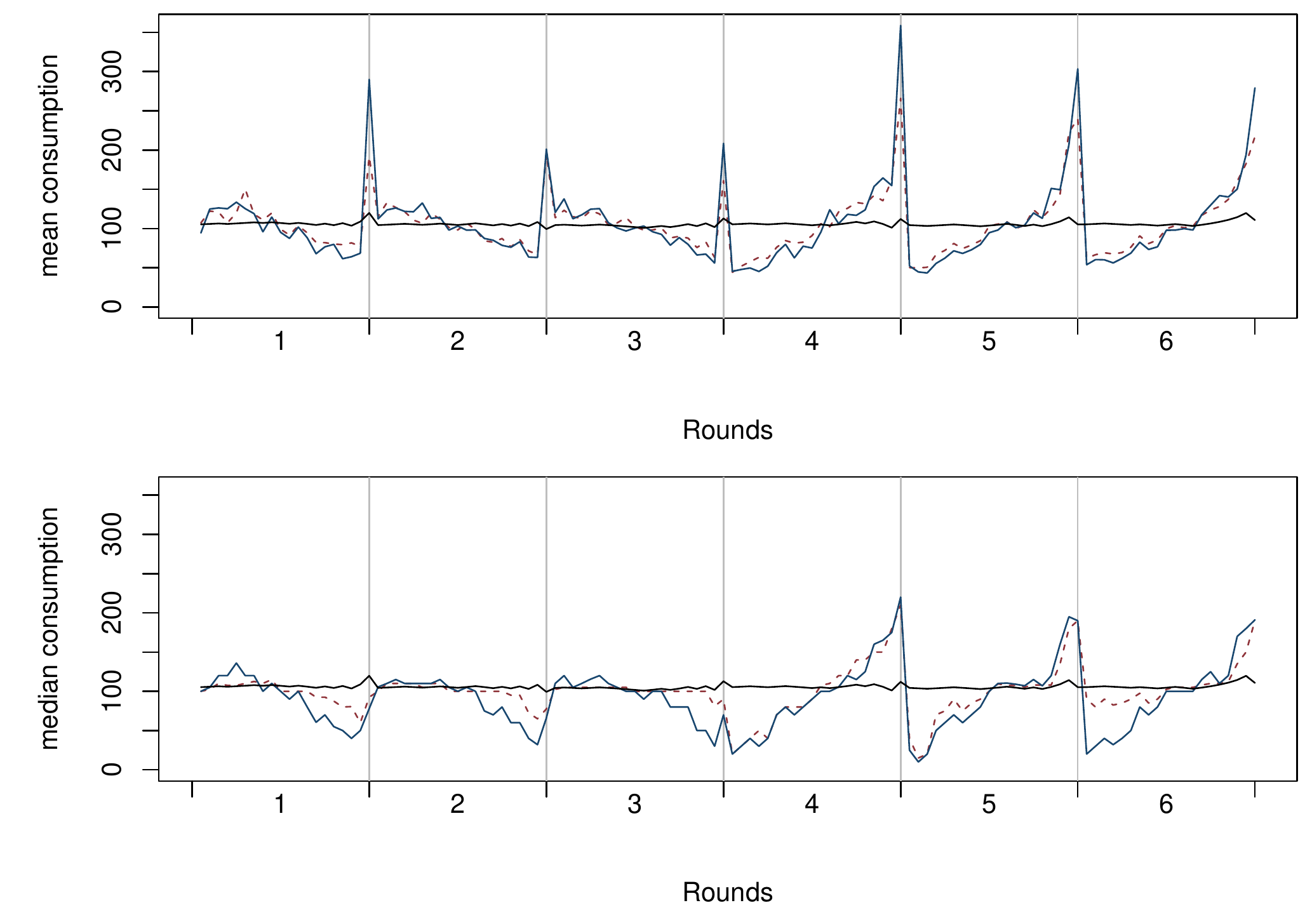}\\
\hspace*{3.7cm}\includegraphics[width=0.6\textwidth]{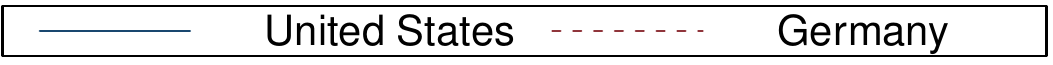}

\end{figure}

Figure \ref{fig:mean_consumption} shows the mean and median consumption of all participants in the Borrowing First sessions (upper two graphs) and the Savings First sessions (lower two graphs). The solid blue lines represent the results for the US sample, while the dashed red lines represent the results for the German sample from \cite{Meissner2016}. The solid black line marks the optimal consumption path according to equations \eqref{eq:opt_solution_1} to \eqref{eq:opt_solution_3}.

For both samples, the mean and median consumption increases steadily over the life-cycle when the income stream is increasing, whereas they decrease steadily over the life-cycle when the income stream is decreasing. Furthermore, in both cases, the mean and median consumption profiles are generally much steeper (i.e. less smooth) for increasing income streams relative to consumption profiles arising from decreasing income streams. Such similarities point towards a comparable behavior of participants across both populations.

To further analyze the individual behavior of participants, we follow \cite{Meissner2016} and define three different ways to measure the deviations from optimal consumption $m_1$, $m_2$, and $m_3$: 
\begin{align}
 m_1 &= \sum_{t=1}^{20} \left(c^\ast_t(w_t)-c_t\right)
 \label{eq:m1} \\ 
 m_2 &= \sum_{t=1}^{20} \left|c^\ast_t(w_t)-c_t\right|  
 \label{eq:m2} \\ 
 m_3 &= \sum_{t=1}^{20} \left(u(c^\ast_t(w_t^\ast))-u(c_t)\right)
 \label{eq:m3},
\end{align}
where $c^\ast_t(w_t)$ is optimal consumption conditional on current wealth and $c^\ast_t(w^\ast_t)$ is the unconditional optimal consumption as a function of the optimal wealth. These measures summarize the accumulated deviations of a participant within each life-cycle, allowing us to study how participants behave under each type of income stream. For example, for $m_1$, any value above zero means that participants are under-consuming, while values below zero imply over-consumption. $m_2$ allows us to measure the absolute deviations from optimal consumption, and $m_3$ the loss of utility derived from such over-/under-consumption

\begin{figure}\begin{center}
\caption{Median aggregate deviations \label{fig:deviations}}
\includegraphics[width=0.9\textwidth]{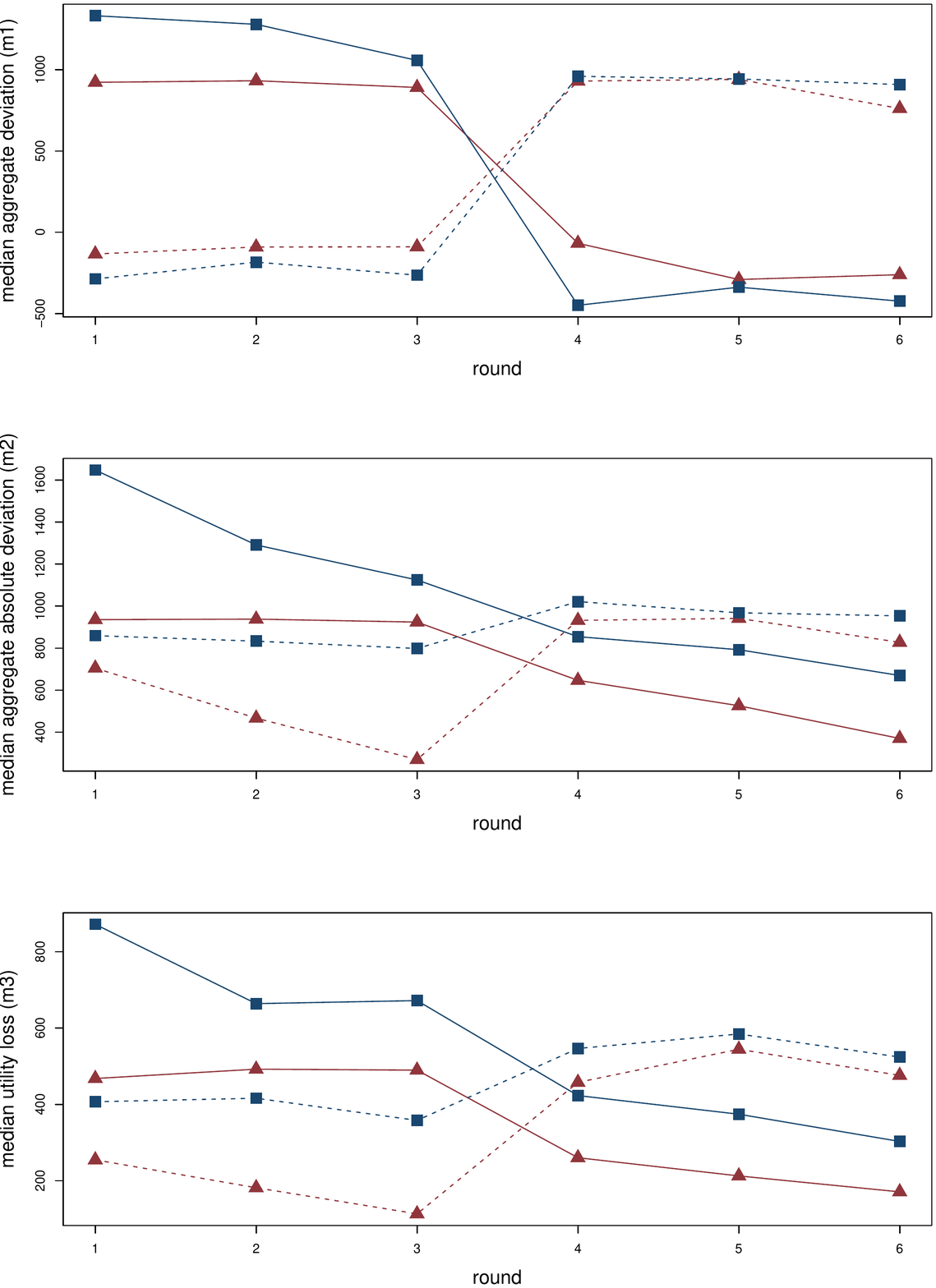}
\includegraphics[width=0.7\textwidth]{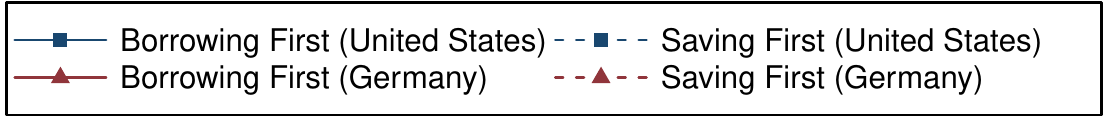}
\end{center}

\end{figure}

In Figure \ref{fig:deviations} we plot the median of $m_1$, $m_2$, and $m_3$ across all participants for each round and country. The behavioral patterns appear to be similar across countries, with both types of participants over-consuming in savings rounds and under-consuming in borrowing rounds (see $m_1$). It is also clear that US participants perform worse than those from Germany, as measure $m_2$ appears to be higher for the US than for Germany (i.e., US participants consume relatively less than Germans in borrowing rounds and consume relatively too much in saving rounds). Importantly, across all three measures, the median deviation is significantly higher for both countries when participants face an increasing income stream (i.e., when they should borrow) than when they face a decreasing income stream (i.e., when they should save).

\begin{table}[t!]
\caption{Median measures $m_1, m_2, m_3$ by country \label{tab:means_medians}}
\begin{center}
\resizebox{\textwidth}{!}{
\begin{tabular}{llD{.}{.}{4.3}D{.}{.}{4.3}D{.}{.}{4.3}D{.}{.}{4.3}D{.}{.}{4.3}D{.}{.}{4.3}}
\toprule
          Measure   & Session & \multicolumn{6}{c}{Round}\\
              &          &\mc{1}  &  \mc{2}  &  \mc{3}  &  \mc{4}   &  \mc{5}  &  \mc{6}\\
\midrule

\multicolumn{8}{c}{{United States}}\\
\midrule
Median $m_1$  & \mc{BF}	  & 1331.968 & 1279.272 & 1057.301 & -448.976 & -337.514 & -423.506\\
              & \mc{SF}	  & -286.689 & -183.513 & -263.750 &  959.476 &  943.169 &  908.636\\[1ex]
$p-$value     &      	  &   <0.001 &   <0.001 &   <0.001 &   <0.001 &   <0.001 &   <0.001\\[1ex]
\midrule
Median $m_2$  & \mc{BF}	  & 1648.105 & 1291.288 & 1124.624 &  854.537 &  792.410 &  669.553\\
              & \mc{SF}	  &  859.351 &  833.197 &  798.277 & 1021.381 &  968.282 &  953.587 \\[1ex]
$p-$value     &      	  &   <0.001 &   <0.001 &   <0.001 &    0.005 &    0.014 &    0.060\\[1ex]
\midrule
Median $m_3$  & \mc{BF}	  &  872.064 &  664.189 &  672.204 &  423.347 &  374.457 &  303.111\\
              & \mc{SF}	  &  407.131 &  416.362 &  358.253 &  546.759 &  584.643 &  524.210\\[1ex]
$p-$value     &      	  &    0.001 &    0.004 &    0.002 &    0.049 &    0.037 &    0.085\\
\midrule
\multicolumn{8}{c}{{Germany}}\\
\midrule
Median $m_1$  & \mc{BF}	  &  922.395 &  932.217 &  890.805 &  -67.671 & -290.294 & -260.731\\
              & \mc{SF}	  & -133.438 &  -90.453 &  -89.256 &  929.985 &  940.529 &  761.430\\[1ex]
$p-$value     &      	  &   <0.001 &   <0.001 &   <0.001 &   <0.001 &   <0.001 &   <0.001\\[1ex]
\midrule
Median $m_2$  & \mc{BF}	  &  935.503 &  938.093 &  923.923 &  646.581 &  525.672 &  369.829\\
              & \mc{SF}	  &  704.668 &  466.493 &  269.223 &  932.037 &  941.523 &  827.583\\[1ex]
$p-$value     &      	  &    0.017 &    0.009 &    0.005 &    0.185 &    0.096 &    0.222\\[1ex]
\midrule
Median $m_3$  & \mc{BF}	  &  468.009 &  492.488 &  489.947 &  260.469 &  212.785 &  171.035\\
              & \mc{SF}	  &  254.897 &  181.869 &  113.385 &  457.959 &  544.636 &  476.155\\[1ex]
$p-$value     &      	  &    0.041 &    0.059 &    0.007 &    0.439 &    0.155 &    0.409\\
\bottomrule
\end{tabular}}

\end{center}
\begin{tablenotes}
\footnotesize
        \item[a] \emph{Notes:} For each country and round we present the median measure ($m_1$ to $m_3$) across participants for each round for each treatment order (BF or SF). The reported p-values are from Mann Whitney U tests comparing the values for each round.
    \end{tablenotes}
\end{table}

In  Table \ref{tab:means_medians}, we report the median of $m_1$, $m_2$, and $m_3$ for each round as well as the p-values from pair-wise Mann-Whitney \textit{U} test comparisons across types of sessions. In most cases, the differences in deviations between treatments are statistically different. Importantly, the relative differences in deviations from optimal consumption in the saving and borrowing rounds are similar across samples. This can be seen in Table \ref{tab:effect size} where we report the effect sizes of the difference in deviations across treatments for each measure and country.\footnote{To report effect sizes we calculate Cohen's D for each measure ($m_1, m_2, m_3$) across both samples. In our case, this is the standardized difference of the mean deviation from optimal consumption between the saving and borrowing treatments. For more details, see \cite{cohen1988statistical}.} In most cases, the effect sizes are relatively close to each other. The exceptions are rounds 1 to 3 for $m_2$, which are slightly larger for the US sample. This difference is most likely driven by the large deviations from optimal consumption in the first round of the savings treatment for the US sample (see the middle panel of Figure \ref{fig:deviations}).

\begin{table}[t]
\caption{Cohen's d in the US and Germany\label{tab:effect size}}
\centering
\begin{tabular}{cccc}
\toprule
measure & country  & Rounds 1-3 & Rounds 4-6  \\
 \midrule
 $m_1$ & US   &  1.310 & 1.156 \\
 & Germany & 1.031  & 1.335  \\[1ex]
$m_2$ & US   & 0.632 & 0.467\\
 & Germany & 0.339 &  0.320\\[1ex]
 $m_3$ & US   & 0.125 & 0.117 \\
 & Germany & 0.146 & 0.268 \\
\bottomrule
\end{tabular}

\end{table}

In fact, while participants in \cite{Meissner2016} seem to improve their consumption decisions over time, the new sample seems to be consistently worse in the borrowing treatment compared to the saving treatment. To analyze the learning of participants, in Table \ref{tab:learning} we replicate Table 2 of \cite{Meissner2016} and present the \textit{median differences} in measure $m_2$ between consecutive rounds ($\Delta_r^{r-1} m2=m_2^{r-1}-m_2^r$) and with the first round ($\Delta_r^{1} m2=m_2^1-m_2^r$).\footnote{Note that differences in the first three rounds compared to the last three rounds may be caused by both learning and the treatment effect.} As in the original experiment, we see that the differences between consecutive rounds of the same treatment (saving or borrowing) are positive and significant in all cases except one. Also, as in \cite{Meissner2016}, participants perform significantly worse in the first round compared to later rounds in BF sessions. However, participants do not perform better in the borrowing rounds compared to the first round in SF sessions. The replication of this result supports the idea that participants perform worse in scenarios requiring borrowing than in scenarios requiring saving and that there is an asymmetric  process in which learning from borrowing rounds spills over to saving rounds, but not the other way around.

\begin{table}[t]
\caption{Learning\label{tab:learning}}
	\begin{center}
	\resizebox{\textwidth}{!}{
		\begin{tabular}{llD{.}{.}{4.3}D{.}{.}{4.3}D{.}{.}{4.3}D{.}{.}{4.3}D{.}{.}{4.3}D{.}{.}{4.3}}
			\toprule
			Measure& Condition & \multicolumn{6}{c}{Round}\\
			&           &\mc{1}  &  \mc{2}  &  \mc{3}  &  \mc{4}   &  \mc{5}  &  \mc{6}\\
			\midrule
			\multicolumn{8}{c}{United States}\\
			\hline
			Median $\Delta_r^{r-1} m2$   & \mc{BF}	  & NA  & 175.824 &  55.160 &  306.502 &  17.784 &   6.571\\
			$p-$value     				 &      	  &     &   0.008 &   0.023 &   <0.001 &   0.061 &   0.455\\[1ex]
			Median $\Delta_r^{1} m2$ 	 & 	      	  & NA  & 175.824 & 398.853 &  737.771 & 890.776 & 883.846\\
			$p-$value     				 &      	  &     &   0.008 &  <0.001 &   <0.001 &  <0.001 &  <0.001\\[1ex]
			\midrule
			Median $\Delta_r^{r-1} m2$   & \mc{SF}	  & NA  & 183.457 &  30.174 & -381.133 &  10.412 &   2.168\\
			$p-$value     				 &      	  &     &   0.001 &   0.102 &   <0.001 &   0.078 &   0.804\\[1ex]
			Median $\Delta_r^{1} m2$     & 	      	  & NA  & 183.457 & 192.474 & -124.496 &  -3.739 & -23.383\\
			$p-$value    	 			 &      	  &     &   0.001 &  <0.001 &    0.286 &   0.741 &   0.757\\[1ex]
		\midrule
			\multicolumn{8}{c}{Germany}\\
			\midrule
			Median $\Delta_r^{r-1} m2$   & \mc{BF}	  & NA  &  58.268 &  18.724 &   69.507 &  98.323 &  19.958\\
			$p-$value     				 &      	  &     &  <0.001 &   0.057 &    0.020 &   0.003 &   0.223\\[1ex]
			Median $\Delta_r^{1} m2$ 	 & 	      	  & NA  &  58.268 & 137.011 &  370.480 & 439.567 & 575.866\\
			$p-$value     				 &      	  &     &  <0.001 &  <0.001 &   <0.001 &  <0.001 &  <0.001\\[1ex]
			\midrule
			Median $\Delta_r^{r-1} m2$   & \mc{SF}	  & NA  &  66.591 &  80.909 & -202.889 &  62.482 &  54.239\\
			$p-$value     				 &      	  &     &  <0.001 &  <0.001 &   <0.001 &   0.143 &   0.007\\[1ex]
			Median $\Delta_r^{1} m2$     & 	      	  & NA  &  66.591 & 155.424 &  -40.365 &  37.363 &  69.948\\
			$p-$value    	 			 &      	  &     &  <0.001 &  <0.001 &    0.413 &   0.752 &   0.381\\[1ex]
			\bottomrule
		\end{tabular}}
	\end{center}
	\begin{tablenotes}
	\footnotesize
        \item[a] \emph{Notes:} For each country and round we present present the \textit{median differences} in measure $m_2$ between consecutive rounds ($\Delta_r^{r-1} m2=m_2^{r-1}-m_2^r$) and with the first round ($\Delta_r^{1} m2=m_2^1-m_2^r$). The reported p-values are from Wilcoxon signed rank tests.
    \end{tablenotes}
\end{table}

\subsubsection{Determinants of Deviations from Optimal Consumption \label{results:deviations}}

To understand what determines deviations from optimal consumption, in Table \ref{tab:m2} we regress the individual $m_2$ for each participant in each round on a series of covariates. In the first column, we use the full sample and include the variable $Germany$ (which takes a value of one for observations from \cite{Meissner2016}) and $Round$, which controls for the round. The results show that German participants tend to have smaller deviations from optimal consumption. This difference in performance is mostly driven by differences in the borrowing rounds as can be seen in Tables \ref{tab:m2saving} and \ref{tab:m2Borrow} of Appendix \ref{sec:additional_tables}, where we reproduce Table \ref{tab:m2} by partitioning the data into saving and borrowing rounds.

\begin{table}[t]
\caption{Determinants of deviations from optimal consumption. \label{tab:m2}}
\centering
{
\def\sym#1{\ifmmode^{#1}\else\(^{#1}\)\fi}
\begin{tabular}{l*{5}{c}}
\toprule
            &\multicolumn{1}{c}{(1)}&\multicolumn{1}{c}{(2)}&\multicolumn{1}{c}{(3)}&\multicolumn{1}{c}{(4)}&\multicolumn{1}{c}{(5)}\\
            &\multicolumn{1}{c}{Combined}&\multicolumn{1}{c}{US}&\multicolumn{1}{c}{US}&\multicolumn{1}{c}{US}&\multicolumn{1}{c}{US}\\
\midrule
Round       &      -64.13\sym{***}&      -76.44\sym{***}&      -71.62\sym{***}&      -71.07\sym{***}&      -71.07\sym{***}\\
            &     (15.40)         &     (23.08)         &     (22.80)         &     (23.94)         &     (24.01)         \\
\addlinespace
Germany     &      -321.5\sym{***}&                     &                     &                     &                     \\
            &     (105.4)         &                     &                     &                     &                     \\
\addlinespace
CRT score   &                     &      -364.2\sym{***}&                     &                     &      -320.4\sym{***}\\
            &                     &     (63.80)         &                     &                     &     (62.88)         \\
\addlinespace
Female      &                     &                     &       528.5\sym{***}&                     &       310.7\sym{**} \\
            &                     &                     &     (154.6)         &                     &     (151.3)         \\
\addlinespace
Risk aversion&                     &                     &                     &       41.63\sym{**} &       17.57         \\
            &                     &                     &                     &     (18.09)         &     (17.83)         \\
\addlinespace
Constant    &      1363.2\sym{***}&      2111.2\sym{***}&      1180.9\sym{***}&      1129.8\sym{***}&      1782.4\sym{***}\\
            &     (87.97)         &     (169.9)         &     (113.8)         &     (178.1)         &     (212.2)         \\
\addlinespace
CRT known   &          No         &         Yes         &          No         &          No         &         Yes         \\
\midrule
\(N\)       &        1002         &         546         &         540         &         510         &         510         \\
adj. \(R^{2}\)&       0.049         &       0.200         &       0.096         &       0.043         &       0.234         \\
\bottomrule
\multicolumn{6}{l}{\footnotesize Standard errors in parentheses}\\
\multicolumn{6}{l}{\footnotesize \sym{*} \(p<0.10\), \sym{**} \(p<0.05\), \sym{***} \(p<0.01\)}\\
\end{tabular}
}

\begin{tablenotes}
\footnotesize
        \item[a] \emph{Notes:} In each column, we regress measure 2 ($m_2$) on different covariates. The first column contains data from Germany and US. Columns (2) to (5) use only data from the US. All standard errors are clustered at the participant level.
    \end{tablenotes}

\end{table}

Additionally, in columns (2) to (5) of Table \ref{tab:m2} we analyze the effect that CRT, gender, and risk aversion have on determining deviations from optimal consumption. These measures were only collected for the US sample, so all analyses on individual characteristics is limited to US participants. In column (2) we analyze the effect of cognitive reflection, using the number of correct answers in the CRT (\emph{CRT Score}). The coefficient is large, negative, and statistically significant, indicating a strong correlation between cognitive reflection ability and deviations from optimal consumption. This is consistent with \cite{Ballinger_etal_2011}, who also report a negative correlation between cognitive ability (albeit measured with a different test) and deviations from optimal consumption.  In columns (3) and (4) we introduce a gender dummy (\emph{Female}) and \emph{Risk Aversion}, which counts the number of safe choices a participant has made in a multiple price list (MPL) risk elicitation task (see Appendix \ref{sec:instructions} for more details). \emph{CRT known} is a control variable that takes the value of one if participants self-reported having seen the CRT previously and zero otherwise. The results show that both females and participants with high risk aversion deviate more from optimal consumption. In column (5) we run the full model, including CRT, gender, and risk aversion. All the results are robust except for risk aversion, which loses explanatory power once we control for CRT and gender.\footnote{In Tables \ref{tab:m2saving} and \ref{tab:m2Borrow} of Appendix \ref{sec:additional_tables} we split our sample into the saving and borrowing treatments, respectively. These tables replicate Table \ref{tab:m2} and show that our results are robust; higher CRT results in better savings and borrowing decisions, while being female and risk aversion results in inferior savings and borrowing decisions in both treatments.}

\subsection{Debt aversion}

Deviations from optimal consumption do not yet imply debt aversion. All else equal, larger debt aversion should lead to larger differences in deviations from optimal behavior between the saving and the borrowing treatment. Therefore, we construct an individual measure of debt aversion by taking the aggregated difference in absolute deviations from conditional optimal consumption (using $m_2$) in the saving and borrowing treatment and normalizing by the aggregated deviations in both treatments. This individual index of debt aversion (\emph{DA}) allows us to compare debt aversion across the two samples and is formally defined as:\footnote{For notational convenience, indices referring to participants are omitted.}  
\begin{equation} 
    DA= \frac{\mathbbm{1}_{BF}\left(\sum_{r=1}^{3}m_2^{r}-\sum_{r=4}^{6}m_2^{r}\right)+ (1-\mathbbm{1}_{BF}) \left( \sum_{r=4}^{6}m_2^{r}-\sum_{r=1}^{3}m_2^{r}\right)}{\sum_{r=1}^{6}m_2^{r}}, 
\end{equation}
where $\mathbbm{1}_{BF}$ is an indicator function that takes the value of one for participants in the Borrowing First sessions and zero otherwise. The larger the debt aversion index, the larger is $m_2$ in rounds that require borrowing relative to those that require savings to consume optimally. The normalization ensures that the measure is limited to the interval $[-1,1]$. A measure of $DA=1$ indicates that a participant only deviates from optimal consumption in the borrowing treatment, and a measure of $DA=-1$ that she only deviates from optimal consumption in the saving treatment. A measure of $DA=0$ indicates that deviations are the same in the borrowing and the saving treatment and thus that there is no debt aversion. Note that this index does not measure debt aversion itself, as is constructed based on deviations from optimal consumption. However, it may serve as a proxy that can be expected to correlate with debt aversion since a more debt averse person will borrow less in the borrowing treatments and therefore have a higher DA in these rounds.\footnote{One might argue that a simpler proxy for debt aversion could be deviations from optimal consumption in the borrowing treatment. We prefer our index, because it controls for other confounding factors. For instance, a person may simply be bad at solving the intertemporal optimization problem, regardless of whether they have to borrow or save. This person would look like they are debt averse according to the deviations in the borrowing treatment only, but not using the debt aversion index.}

\begin{figure}[t]
\caption{Debt Aversion in Germany and the US \label{fig:KdensityGERUSA}}
\includegraphics[width=\textwidth]{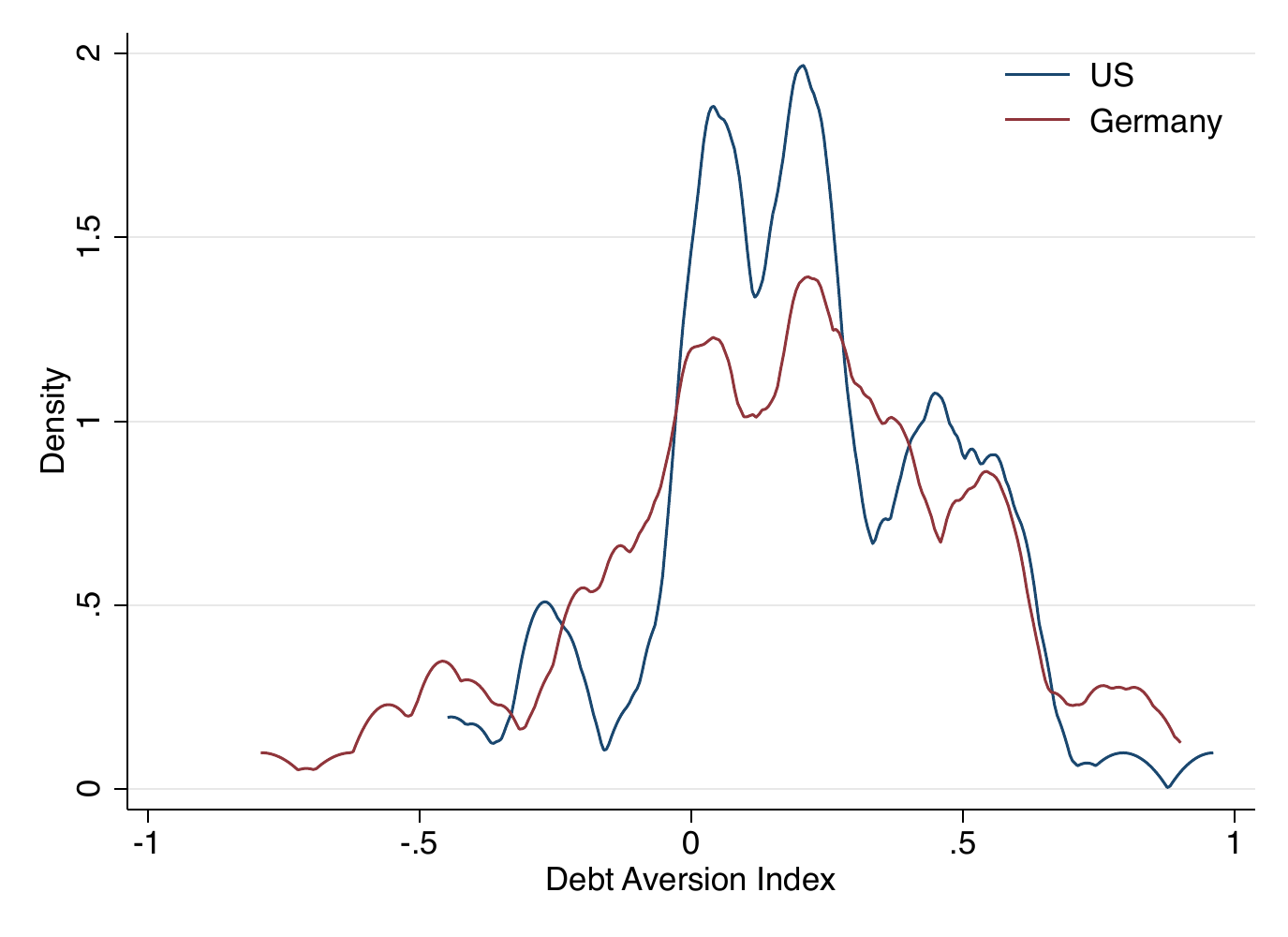}

\end{figure}

Figure \ref{fig:KdensityGERUSA} illustrates the distribution of the debt aversion index in Germany and the US. A Mann-Whitney U test fails to reject a difference in distributions between the German and the US data ($p= 0.5644$).

\begin{table}[t!]
\footnotesize
\caption{Individual characteristics and debt aversion\label{tab:DA}}
\centering
{
\def\sym#1{\ifmmode^{#1}\else\(^{#1}\)\fi}
\begin{tabular}{l*{7}{c}}
\toprule
            &\multicolumn{1}{c}{(1)}&\multicolumn{1}{c}{(2)}&\multicolumn{1}{c}{(3)}&\multicolumn{1}{c}{(4)}&\multicolumn{1}{c}{(5)}&\multicolumn{1}{c}{(6)}&\multicolumn{1}{c}{(7)}\\
            &\multicolumn{1}{c}{Combined}&\multicolumn{1}{c}{US}&\multicolumn{1}{c}{US}&\multicolumn{1}{c}{US}&\multicolumn{1}{c}{US}&\multicolumn{1}{c}{US}&\multicolumn{1}{c}{US}\\
\midrule
Saving First&      -0.215\sym{***}&      -0.252\sym{***}&      -0.242\sym{***}&      -0.241\sym{***}&      -0.247\sym{***}&      -0.258\sym{***}&      -0.260\sym{***}\\
            &     (0.044)         &     (0.050)         &     (0.050)         &     (0.054)         &     (0.054)         &     (0.054)         &     (0.054)         \\
\addlinespace
Germany     &      -0.044         &                     &                     &                     &                     &                     &                     \\
            &     (0.045)         &                     &                     &                     &                     &                     &                     \\
\addlinespace
CRT score   &                     &       0.040\sym{*}  &                     &                     &       0.038         &       0.179\sym{*}  &                     \\
            &                     &     (0.023)         &                     &                     &     (0.025)         &     (0.092)         &                     \\
\addlinespace
Female      &                     &                     &      -0.043         &                     &      -0.006         &       0.007         &       0.003         \\
            &                     &                     &     (0.052)         &                     &     (0.059)         &     (0.059)         &     (0.060)         \\
\addlinespace
Risk aversion&                     &                     &                     &       0.000         &       0.001         &      -0.000         &       0.001         \\
            &                     &                     &                     &     (0.007)         &     (0.008)         &     (0.008)         &     (0.008)         \\
\addlinespace
CRT score\(^{2}\)&                     &                     &                     &                     &                     &      -0.044         &                     \\
            &                     &                     &                     &                     &                     &     (0.028)         &                     \\
\addlinespace
CRT1        &                     &                     &                     &                     &                     &                     &       0.094         \\
            &                     &                     &                     &                     &                     &                     &     (0.101)         \\
\addlinespace
CRT2        &                     &                     &                     &                     &                     &                     &       0.187\sym{**} \\
            &                     &                     &                     &                     &                     &                     &     (0.084)         \\
\addlinespace
CRT3        &                     &                     &                     &                     &                     &                     &       0.129         \\
            &                     &                     &                     &                     &                     &                     &     (0.080)         \\
\addlinespace
Constant    &       0.318\sym{***}&       0.250\sym{***}&       0.346\sym{***}&       0.324\sym{***}&       0.242\sym{***}&       0.197\sym{**} &       0.205\sym{**} \\
            &     (0.038)         &     (0.058)         &     (0.042)         &     (0.065)         &     (0.085)         &     (0.089)         &     (0.091)         \\
\addlinespace
CRT known   &          No         &         Yes         &          No         &          No         &         Yes         &         Yes         &         Yes         \\
\midrule
\(N\)       &         167         &          91         &          90         &          85         &          85         &          85         &          85         \\
adj. \(R^{2}\)&       0.118         &       0.218         &       0.194         &       0.185         &       0.184         &       0.199         &       0.192         \\
\bottomrule
\multicolumn{8}{l}{\footnotesize Standard errors in parentheses}\\
\multicolumn{8}{l}{\footnotesize \sym{*} \(p<0.10\), \sym{**} \(p<0.05\), \sym{***} \(p<0.01\)}\\
\end{tabular}
}

\footnotesize
\begin{tablenotes}
        \item[a] \emph{Notes:} In each column, we regress the debt aversion index (\emph{DA}) on different covariates. The first column contains data from Germany and US. Columns (2) to (7) use only data from the US. CRT1, CRT2 and CRT3 are dummy variables that take the value of one for participants who correctly answer one, two or three CRT questions respectively.
    \end{tablenotes}
\end{table}

Table \ref{tab:DA} contains regressions where the index of debt aversion (\emph{DA}) is the dependent variable. In specification (1) we use the combined data of the US and Germany and control for country and order effects. \emph{Saving First} is a treatment dummy that takes the value of one for participants in the Saving First sessions, while \textit{Germany} is a dummy that takes the value of one if the observation belongs to the original German sample. The results show that participants who start with the saving treatment are less debt averse. However, this is likely an artifact caused by learning effects. As shown in Table \ref{tab:learning} and Figure \ref{fig:deviations}, learning from saving rounds spills over to borrowing rounds, but learning from borrowing rounds has a smaller impact on behavior in the saving rounds. This asymmetry in learning spillovers results in lower perceived \emph{DA} for those participants in SF sessions. 

Importantly, in specification (1) we detect no differences across countries. While the coefficient for the country dummy is negative, which would indicate that German students are less debt averse than those from the US, the effect is small and not statistically significant. This result implies that there are no systematic differences between the levels of debt aversions between American and German students and, therefore, that the original results of \cite{Meissner2016} are robust to different credit cultures and (likely) experience acquiring debt. 

In specifications (2) to (5) we only include observations from US participants to study the effect of different covariates on \emph{DA}.\footnote{One could be worried about multicollinearity, as gender and risk aversion are typically found to be correlated. In our data females are more risk averse ($\rho=0.319$, $p=0.003$) and have lower CRT scores ($\rho=-0.2736$, $p=0.009$). However, the variance inflation factors are no larger than 1.16 for any of the included variables, which indicates that multicollinearity is not a problem.} Specifications (2) to (4) show that only CRT has a weak positive correlation with debt aversion: participants with higher CRT scores appear to be more debt averse. Gender and risk aversion do not seem to be correlated with debt aversion. However, after controlling for gender and risk aversion in specification (5), CRT appears to lose explanatory power ($p=0.13$). Upon closer inspection, it seems that the effect of CRT on debt aversion could be non-linear. In specifications (6) and (7) we include a squared term on the CRT score and dummy variables corresponding to each possible number of correct answers in the CRT respectively. The latter specification shows that the coefficients on all CRT dummies are positive and that answering two questions correctly has a significant positive impact on debt aversion. The dummy on answering three questions correctly is smaller than that of two, and is not statistically significant ($p=0.11$), consistent with a potential non-linear relationship. Including a squared CRT score term in specifications (6), yields a negative squared term and CRT becomes marginally significant ($p=0.055$), even after controlling for gender and risk aversion. In summary, there seems to be some weak evidence for a positive correlation between CRT and debt aversion. After accounting for a potential non-linear relationship, this evidence becomes somewhat stronger. Evidence for a positive correlation between CRT and debt aversion would be interesting, as CRT has the opposite effect on deviations from optimal consumption (see Section \ref{results:deviations}). As our debt aversion index is built using deviations from optimal behavior, this suggests that participants with a higher CRT score generally deviate less from optimal consumption, but have a higher asymmetry in deviations from optimal consumption in the borrowing and saving condition, compared to participants with lower CRT score. 
However, given the weak association, we would caution against over-interpreting this result.

\FloatBarrier

\section{Conclusion}\label{sec:conclusion}

\cite{Meissner2016} runs a life-cycle consumption and saving experiment in which he shows that participants perform relatively worse when they need to borrow to consume optimally than when they need to save. This asymmetry is interpreted as a tendency to avoid getting in debt, that is: debt aversion. However, participants in the original experiment are undergraduate students from a large public university in Germany. Therefore, it is possible that the observed debt aversion in \cite{Meissner2016} is limited to the specific population it considers. Germany is known for its low debt levels and for a tradition of shunning debt. Moreover, undergraduate students of public universities in Germany are unlikely to have any experience acquiring debt, which might also contribute to \cite{Meissner2016}'s results \citep{Duffy2016}. 

The present paper replicates \cite{Meissner2016} with undergraduate students from the United States. The United States is known to be more tolerant towards debt \citep{calder2009financing} and to encourage it through its institutions \citep{garon2011beyond}. All of the main findings from the original study replicate with similar effect sizes, confirming the importance of debt aversion even within a population that is likely more exposed to debt. Importantly, we do not find evidence suggesting that debt aversion differs between participants from the US and Germany. Additionally, we extend \cite{Meissner2016} by constructing an individual measure of debt aversion and testing whether it correlates with individual characteristics of our participants. We do not detect any effect of gender or risk preferences on the levels of debt aversion. Interestingly, we find that the CRT score is negatively correlated with deviations from optimal consumption, but weakly positively correlated with debt aversion. However,  we would caution against overinterpreting this result, as the evidence is rather weak.  

To conclude, our paper contributes by successfully replicating a pioneering experiment on debt aversion. We do so by using a population that \textit{a priori} could be expected to have a more positive attitude towards debt and more experience using it. Nonetheless all of the main findings are replicated. Additionally, we extend the original paper by analyzing how the individual characteristics of participants affect their debt aversion.

\clearpage
\FloatBarrier

\bibliography{BibliographyTEMP}
\bibliographystyle{chicago}
\clearpage
\begin{appendices}

\section{Additional tables}

\label{sec:additional_tables}
\subsection{Borrowing/Saving sample split}
\begin{table}[h!]
\caption{Measure 2 ($m_2$) - Saving treatment only \label{tab:m2saving}} 

\centering
{
\def\sym#1{\ifmmode^{#1}\else\(^{#1}\)\fi}
\begin{tabular}{l*{5}{c}}
\toprule
            &\multicolumn{1}{c}{(1)}&\multicolumn{1}{c}{(2)}&\multicolumn{1}{c}{(3)}&\multicolumn{1}{c}{(4)}&\multicolumn{1}{c}{(5)}\\
            &\multicolumn{1}{c}{Combined}&\multicolumn{1}{c}{US}&\multicolumn{1}{c}{US}&\multicolumn{1}{c}{US}&\multicolumn{1}{c}{US}\\
\midrule
Round       &      -53.97\sym{*}  &      -71.93\sym{**} &      -67.10\sym{*}  &      -71.87\sym{*}  &      -80.15\sym{**} \\
            &     (29.83)         &     (33.85)         &     (38.50)         &     (41.85)         &     (36.23)         \\
\addlinespace
Germany     &      -209.7\sym{*}  &                     &                     &                     &                     \\
            &     (109.1)         &                     &                     &                     &                     \\
\addlinespace
CRT score   &                     &      -367.2\sym{***}&                     &                     &      -335.2\sym{***}\\
            &                     &     (70.87)         &                     &                     &     (70.35)         \\
\addlinespace
Female      &                     &                     &       459.9\sym{***}&                     &       238.1\sym{*}  \\
            &                     &                     &     (154.5)         &                     &     (139.5)         \\
\addlinespace
Risk aversion&                     &                     &                     &       29.31\sym{*}  &       8.614         \\
            &                     &                     &                     &     (16.30)         &     (15.15)         \\
\addlinespace
Constant    &      1090.9\sym{***}&      1872.7\sym{***}&       959.8\sym{***}&       978.3\sym{***}&      1698.5\sym{***}\\
            &     (132.9)         &     (225.5)         &     (174.4)         &     (211.0)         &     (254.1)         \\
\addlinespace
CRT known   &          No         &         Yes         &          No         &          No         &         Yes         \\
\midrule
\(N\)       &         501         &         273         &         270         &         255         &         255         \\
adj. \(R^{2}\)&       0.031         &       0.280         &       0.099         &       0.030         &       0.299         \\
\bottomrule
\multicolumn{6}{l}{\footnotesize Standard errors in parentheses}\\
\multicolumn{6}{l}{\footnotesize \sym{*} \(p<0.10\), \sym{**} \(p<0.05\), \sym{***} \(p<0.01\)}\\
\end{tabular}
}

\begin{tablenotes}
\footnotesize
        \item[a] \emph{Notes:} In each column, we regress measure 2 ($m_2$) on different covariates. The first column contains data from Germany and US. Columns (2) to (5) use only data from the US. All standard errors are clustered at the participant level.
    \end{tablenotes}
\end{table} 

\begin{table}[h!]
\caption{Measure 2 ($m_2$) - Borrowing treatment only \label{tab:m2Borrow}} 
\centering
{
\def\sym#1{\ifmmode^{#1}\else\(^{#1}\)\fi}
\begin{tabular}{l*{5}{c}}
\toprule
            &\multicolumn{1}{c}{(1)}&\multicolumn{1}{c}{(2)}&\multicolumn{1}{c}{(3)}&\multicolumn{1}{c}{(4)}&\multicolumn{1}{c}{(5)}\\
            &\multicolumn{1}{c}{Combined}&\multicolumn{1}{c}{US}&\multicolumn{1}{c}{US}&\multicolumn{1}{c}{US}&\multicolumn{1}{c}{US}\\
\midrule
Round       &      -78.68\sym{**} &      -90.28\sym{**} &      -84.81\sym{*}  &      -74.63         &      -67.49         \\
            &     (32.58)         &     (43.36)         &     (44.57)         &     (50.12)         &     (46.37)         \\
\addlinespace
Germany     &      -434.6\sym{***}&                     &                     &                     &                     \\
            &     (118.9)         &                     &                     &                     &                     \\
\addlinespace
CRT score   &                     &      -359.8\sym{***}&                     &                     &      -306.2\sym{***}\\
            &                     &     (71.76)         &                     &                     &     (72.98)         \\
\addlinespace
Female      &                     &                     &       593.9\sym{***}&                     &       382.8\sym{*}  \\
            &                     &                     &     (179.7)         &                     &     (195.5)         \\
\addlinespace
Risk aversion&                     &                     &                     &       53.73\sym{**} &       27.58         \\
            &                     &                     &                     &     (24.27)         &     (26.09)         \\
\addlinespace
Constant    &      1652.0\sym{***}&      2380.1\sym{***}&      1434.7\sym{***}&      1298.2\sym{***}&      1879.5\sym{***}\\
            &     (125.9)         &     (196.8)         &     (170.1)         &     (264.1)         &     (277.4)         \\
\addlinespace
CRT known   &          No         &         Yes         &          No         &          No         &         Yes         \\
\midrule
\(N\)       &         501         &         273         &         270         &         255         &         255         \\
adj. \(R^{2}\)&       0.073         &       0.175         &       0.108         &       0.059         &       0.222         \\
\bottomrule
\multicolumn{6}{l}{\footnotesize Standard errors in parentheses}\\
\multicolumn{6}{l}{\footnotesize \sym{*} \(p<0.10\), \sym{**} \(p<0.05\), \sym{***} \(p<0.01\)}\\
\end{tabular}
}

\begin{tablenotes}
\footnotesize
        \item[a] \emph{Notes:} In each column, we regress measure 2 ($m_2$) on different covariates. The first column contains data from Germany and US. Columns (2) to (5) use only data from the US. All standard errors are clustered at the participant level.
    \end{tablenotes}
\end{table}

\FloatBarrier
\clearpage

\section{Instructions}
\label{sec:instructions}

\includepdf[pages={1,2,3,4,5,6,7,8,9,10}]{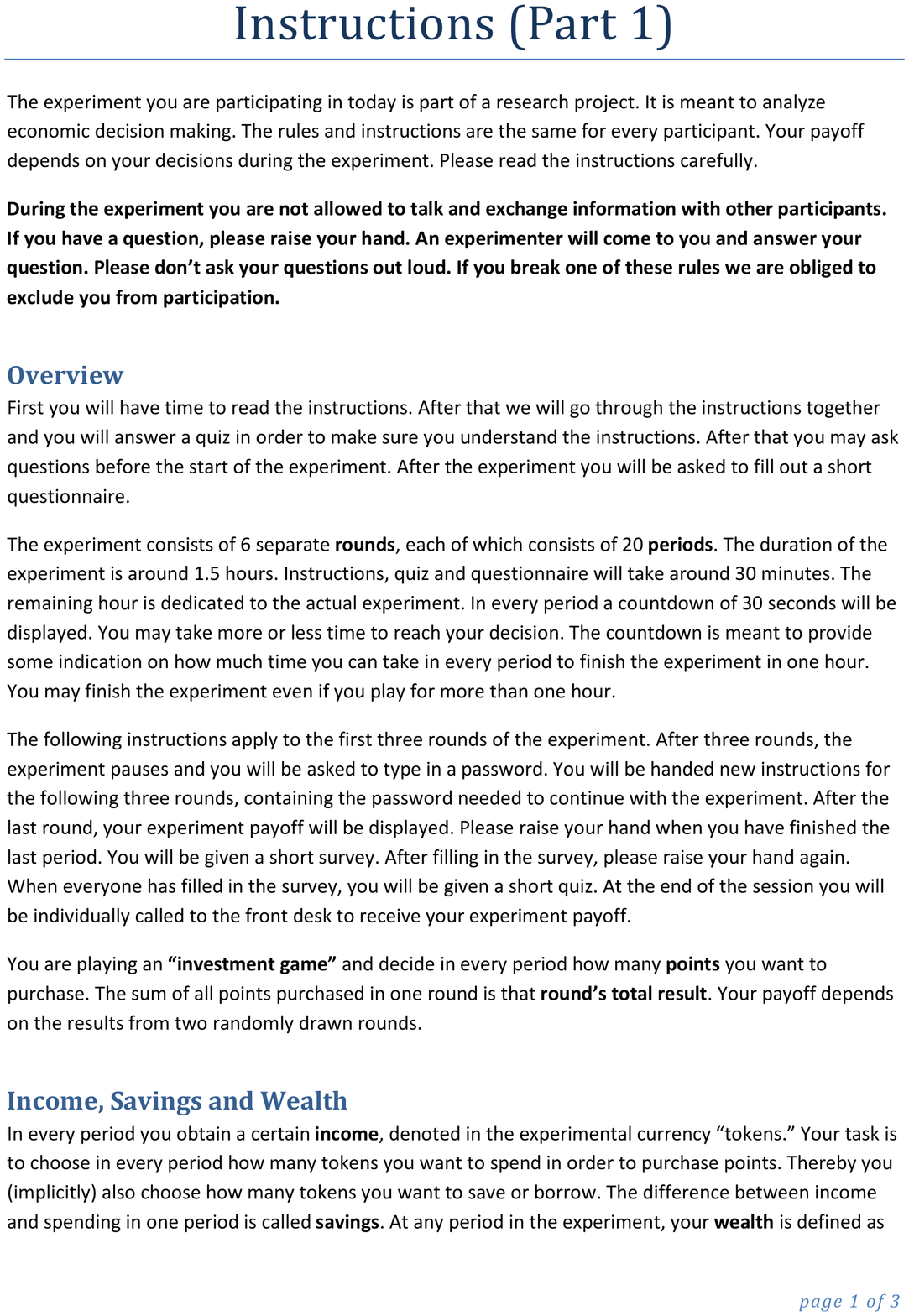}

\end{appendices}

\end{document}